\documentclass[iop, twocolumn, numberedappendix, tighten]{emulateapj}

\usepackage{datetime}
\usepackage{graphicx}
\usepackage{dcolumn}
\usepackage{times}
\usepackage{amsmath}
\usepackage{enumerate}
\usepackage{natbib}
\usepackage{url}
\usepackage[utf8]{inputenc}
\usepackage{color}
\usepackage[colorlinks,urlcolor=blue,citecolor=cyan,linkcolor=red]{hyperref}
\usepackage{microtype}

\newcommand{\beq}{\begin{equation}}
\newcommand{\eeq}{\end{equation}}

\def\EAH{\textit{Einstein@Home}}
\def\psr{{PSR~J1950$+$2414}}
\def\ttwo{\texttt{TEMPO2}}

\submitted{To appear in ApJ}

\shorttitle{\EAH{} Discovery of a Millisecond Pulsar in an Eccentric Binary Orbit}
\shortauthors{Knispel et al.}

\begin{document}

\title{\EAH{} Discovery of a PALFA Millisecond Pulsar in an Eccentric Binary Orbit}

\author{
B.~Knispel\altaffilmark{1,2,$\dagger$},
A.~G.~Lyne\altaffilmark{3},
B.~W.~Stappers\altaffilmark{3},
P.~C.~C.~Freire\altaffilmark{4},
P.~Lazarus\altaffilmark{4},
B.~Allen\altaffilmark{2,5,1},
C.~Aulbert\altaffilmark{2},
O.~Bock\altaffilmark{2},
S.~Bogdanov\altaffilmark{6},
A.~Brazier\altaffilmark{7,8},
F.~Camilo\altaffilmark{6},
F.~Cardoso\altaffilmark{9},
S.~Chatterjee\altaffilmark{7},
J.~M.~Cordes\altaffilmark{7},
F.~Crawford\altaffilmark{10},
J.~S.~Deneva\altaffilmark{11},
H.-B.~Eggenstein\altaffilmark{2},
H.~Fehrmann\altaffilmark{2},
R.~Ferdman\altaffilmark{12},
J.~W.~T.~Hessels\altaffilmark{13,14},
F.~A.~Jenet\altaffilmark{15},
C.~Karako-Argaman\altaffilmark{12},
V.~M.~Kaspi\altaffilmark{12},
J.~van~Leeuwen\altaffilmark{13,14}, 
D.~R.~Lorimer\altaffilmark{9},
R.~Lynch\altaffilmark{12},
B.~Machenschalk\altaffilmark{2},
E.~Madsen\altaffilmark{12},
M.~A.~McLaughlin\altaffilmark{9},
C.~Patel\altaffilmark{12},
S.~M.~Ransom\altaffilmark{16},
P.~Scholz\altaffilmark{12},
X.~Siemens\altaffilmark{5},
L.~G.~Spitler\altaffilmark{4},
I.~H.~Stairs\altaffilmark{17},
K.~Stovall\altaffilmark{18},
J.~K.~Swiggum\altaffilmark{9},
A.~Venkataraman\altaffilmark{19},
R.~S.~Wharton\altaffilmark{7},
W.~W.~Zhu\altaffilmark{4,17}
}
\altaffiltext{$\dagger$}{Email: \href{mailto:benjamin.knispel@aei.mpg.de}{benjamin.knispel@aei.mpg.de}}
\altaffiltext{1}{Leibniz Universit{\"a}t, Hannover, D-30167 Hannover, Germany}
\altaffiltext{2}{Max-Planck-Institut f{\"u}r Gravitationsphysik, Callinstr.\ 38, D-30167 Hannover, Germany}
\altaffiltext{3}{Jodrell Bank Centre for Astrophysics, School of Physics and Astronomy, University of Manchester, Manchester, M13 9PL, UK}
\altaffiltext{4}{Max-Planck-Institut f{\"u}r Radioastronomie, Auf dem H{\"u}gel 69, 53121 Bonn, Germany}
\altaffiltext{5}{Physics Department, University of Wisconsin -- Milwaukee, Milwaukee WI 53211, USA}
\altaffiltext{6}{Columbia Astrophysics Laboratory, Columbia University, New York, NY 10027, USA}
\altaffiltext{7}{Department of Astronomy and Center for Radiophysics and Space Research, Cornell University, Ithaca, NY 14853, USA}
\altaffiltext{8}{Cornell Center for Advanced Computing, Rhodes Hall, Cornell, University, Ithaca, NY 14853, USA}
\altaffiltext{9}{Department of Physics and Astronomy, West Virginia University, Morgantown, WV 26506, USA}
\altaffiltext{10}{Department of Physics and Astronomy, Franklin and Marshall College, Lancaster, PA 17604-3003, USA} 
\altaffiltext{11}{National Academies of Science, resident at the Naval Research Laboratory, Washington, DC 20375, USA}
\altaffiltext{12}{Department of Physics, McGill University, Montreal, QC H3A 2T8, Canada} 
\altaffiltext{13}{ASTRON, Netherlands Institute for Radio Astronomy, Postbus 2, 7990 AA, Dwingeloo, The Netherlands} 
\altaffiltext{14}{Anton Pannekoek Institute for Astronomy, University of Amsterdam, Science Park 904, 1098 XH Amsterdam, The Netherlands}
\altaffiltext{15}{Center for Gravitational Wave Astronomy, University of Texas at Brownsville, TX 78520, USA}
\altaffiltext{16}{NRAO, Charlottesville, VA 22903, USA} 
\altaffiltext{17}{Department of Physics and Astronomy, University of British Columbia, 6224 Agricultural Road Vancouver, BC V6T 1Z1, Canada}
\altaffiltext{18}{Department of Physics and Astronomy, University of New Mexico, NM, 87131, USA}
\altaffiltext{19}{Arecibo Observatory, HC3 Box 53995, Arecibo, PR 00612, USA}

\begin{abstract}
We report the discovery of the millisecond pulsar (MSP) \psr{}
($P=4.3$\,ms) in a binary system with an eccentric ($e=0.08$) 22-day
orbit in Pulsar ALFA survey observations with the Arecibo
telescope. Its companion star has a median mass of 0.3\,$M_\odot$ and
is most likely a white dwarf. Fully recycled MSPs like this one are
thought to be old neutron stars spun-up by mass transfer from a
companion star. This process should circularize the orbit, as is
observed for the vast majority of binary MSPs, which predominantly
have orbital eccentricities $e < 0.001$. However, four recently
discovered binary MSPs have orbits with $0.027 < e < 0.44$; \psr{} is
  the fifth such system to be discovered. The upper limits for its
  intrinsic spin period derivative and inferred surface magnetic field
  strength are comparable to those of the general MSP population. The
  large eccentricities are incompatible with the predictions of the
  standard recycling scenario: something unusual happened during their
  evolution. Proposed scenarios are a) initial evolution of the pulsar
  in a triple system which became dynamically unstable, b) origin in
  an exchange encounter in an environment with high stellar density,
  c) rotationally delayed accretion-induced collapse of a
  super-Chandrasekhar white dwarf, and d) dynamical interaction of the
  binary with a circumbinary disk. We compare the properties of all
  five known eccentric MSPs with the predictions of these formation
  channels. Future measurements of the masses and proper motion might
  allow us to firmly exclude some of the proposed formation scenarios.
\end{abstract}


\section{Introduction}
\label{sec:introduction}

Millisecond pulsars (MSPs; \citealp{1982Natur.300..615B}) are thought
to be old neutron stars (NS) spun up by mass accretion and transfer of
angular momentum from a companion star \citep{1982CSci...51.1096R,
  1982Natur.300..728A}. The spin frequencies of these so-called
``recycled'' pulsars can range up to 716\,Hz
(\citealp{2006Sci...311.1901H}; ATNF pulsar catalog,
\citealp{2005AJ....129.1993M}). The different evolutionary phases of
these pulsars in binary systems and also the transitions between these
phases have been recently observed in much more detail, though several
major puzzles still remain. The binary system starts off as a low-mass
X-ray binary in which a NS accretes matter from a companion star
\citep{1976ApJ...207..574S, 1997ApJS..113..367B}. The main emission
from these system is X-rays from the hot accretion disk. These system
can transition into an accreting X-ray MSP in a binary
\citep{1998Natur.394..344W}, in which matter is funnelled onto the
neutron star's surface and significant X-ray emission modulated by the
NS spin is detected. After the accretion dies off, the NS can become
``visible'' as a radio MSP, powered by the rotation of the neutron
star's magnetic field \citep{1999A&A...350..928T,
  2004Sci...304..547S}. In some cases, these systems are seen to
switch on roughly year-long timescales between states as an LMXB and a
radio MSP. For example, \citet{2009Sci...324.1411A} showed that the
radio MSP PSR~J1023$+$0038 has turned on after a recent
($\sim$10\,yrs) LMXB phase; more recently this system changed back
into an LMXB \citep{2014arXiv1412.1306A, 2014ApJ...781L...3P,
  2014ApJ...790...39S}.  Similarly, LMXB/radio MSP state transitions
have been shown for PSRs J1824$-$2452 and J1227$-$4853
\citep{2013Natur.501..517P, 2014MNRAS.441.1825B, 2015ApJ...800L..12R}.
Together, these observations nicely demonstrate the {\it basic}
recycling scenario; however, they have also raised new puzzles and
shown that the details of the process are quite complex. It is also
possible that these three aforementioned transitional MSP systems, all
``redback'' MSPs \citep{2013IAUS..291..127R}, are not representative
of the evolution of all types of MSPs and that these systems might not
lead to recycled pulsars at all \citep{2013ApJ...775...27C}.  As such,
much remains to be understood in the formation of radio MSPs.

The recycling pathway results in highly circular orbits of the binary
system through tidal forces acting during the $10^8 - 10^9$ year long
accretion phase \citep{1994ARA&A..32..591P}.  Until 2008, this seemed
true of all fully recycled MSPs. All binary pulsars with
$\nu\geq50$\,Hz and $\dot\nu\leq10^{-14}$\,Hz\,s$^{-1}$ outside of
globular clusters had orbital eccentricities between $e=10^{-7}$ and
$e=10^{-3}$ \citep{2005AJ....129.1993M}.These eccentricity limits do
not apply to MSPs in globular clusters because their high stellar
densities and resultant close stellar encounters can significantly
increase the orbital eccentricity of a binary MSP after the end of the
accretion phase \citep{1995ApJ...445L.133R, 1996MNRAS.282.1064H} or
even in some cases lead to exchange encounters. Indeed, several highly
eccentric binary MSPs have been found in the Galactic globular cluster
system (e.g., \citet{2005Sci...307..892R}; see also the online catalog
of globular cluster pulsars at
\url{http://www.naic.edu/~pfreire/GCpsr.html}).

In 2008, the situation changed with the discovery of
PSR~J1903$+$0327. This fast-spinning MSP ($P = 2.15$\,ms) in an
eccentric orbit ($e=0.44$) with a $\sim1\,M_{\odot}$ main-sequence
star \citep{2008Sci...320.1309C} cannot have formed through the
``normal'' binary evolution described above. Rather, it is believed to
have originated from a hierarchical triple that became dynamically
unstable \citep{2011MNRAS.412.2763F, 2011ApJ...734...55P,
  2012MNRAS.424.2914P}.

Soon afterwards, \citet{2010NewAR..54...80B} reported an ``anomalous''
orbital eccentricity of 0.027 for PSR~J1618$-$3921 (Bailes; private
communication), a MSP first reported in
\citet{2001ApJ...553..801E}. More recently,
\citet{2013ApJ...775...51D} and \citet{2013MNRAS.435.2234B} reported
the discovery of two more unusual binary MSP systems PSR~J2234$+$06
($e = 0.13$) and PSR~J1946$+$3417 ($e = 0.14$). These three systems
are fully recycled with spin periods between 3 and 12\,ms, orbital
periods $P_\text{orb}$ from 22 to 32\,days and median companion masses
$M_2\approx 0.25\,M_\odot$; i.e., apart from the large orbital
eccentricity all parameters are compatible with the canonical
recycling formation channel leading to a MSP with a white dwarf (WD)
companion as described above.

These unusual orbital eccentricities require a non-standard formation
channel. Like PSR~J1903+0327, they \textit{could} have formed in a
triple system which later became unstable and ejected the outer
(tertiary) companion. In fact, fully recycled MSPs can form in stable
triple systems, as shown by the discovery of PSR~J0337$+$1715, the
first MSP in a stellar triple system with two white dwarf companions
\citep{2014Natur.505..520R, 2014ApJ...781L..13T}.

However, the chaotic disruption of a triple system would most likely
not lead to the formation of MSPs with very similar orbital and spin
characteristics as discussed above. These similarities suggest instead
a more orderly mechanism with a more predictable outcome.

A possibility for such a mechanism was proposed by
\citet{2014MNRAS.438L..86F}.  They suggest these systems formed
through an accretion-induced collapse (AIC) of a super-Chandrasekhar
mass oxygen-neon-magnesium white dwarf in a close binary. This star
initially avoids AIC due to its rapid rotation. Only after the end of
the accretion episode, and after the WD loses sufficient spin angular
momentum, does it undergo AIC to directly produce an MSP in an
eccentric orbit. A second possibility was suggested by
\citet{2014ApJ...797L..24A}: here the MSPs form through the usual
channel, but the orbital eccentricity arises from the dynamical
interaction with a circumbinary disk. This disk may form from donor
material ejected during hydrogen-shell flash
episodes. \citet{2014ApJ...797L..24A} shows that even a short-lived
disk can produce eccentricities as large as $e=0.15$.

Here we present the discovery and initial timing of the binary MSP
\psr{} in Pulsar ALFA (PALFA) survey data obtained at 1.4\,GHz with
the Arecibo telescope. This is the fifth eccentric MSP in the Galactic
field to be discovered; its orbital and spin parameters are similar to
those of PSR~J2234$+$06 and PSR~J1946$+$3417.

First, we will briefly describe the PALFA survey, the \EAH{} project
and its analysis of PALFA survey data, and the discovery of \psr{}. We
then describe the timing observations, data reduction, and timing
solution, followed by a discussion of \psr{}. We discuss our discovery
in the context of different possible formation channels for MSPs in
eccentric binaries. We conclude with an overview of future studies of
this pulsar system and how they might allow us to exclude some of the
possible formation channels.

\subsection{The PALFA Survey}\label{subsec:palfasurvey}
The PALFA Survey \citep{2006ApJ...637..446C} was proposed and is
managed by the PALFA Consortium. It consists of about 40 researchers
(including students) at about ten institutions
worldwide\footnote{\url{http://www2.naic.edu/alfa/pulsar/}}.

At Arecibo Observatory, the PALFA Consortium uses the Arecibo L-band
Feed Array (ALFA\footnote{\url{http://www.naic.edu/alfa/}}). The
output of the seven ALFA beams is fed into the Mock
spectrometers\footnote{Details of the Mock spectrometers may be found
  on the following NAIC web page:
  \url{http://www.naic.edu/~phil/hardware/pdev/pdev.html}}.  The
observing band of 322.6\,MHz is split into two overlapping bands with
bandwidths of 172.0625\,MHz each. The two sub-bands are centred at
1300.1680\,MHz and 1450.1680\,MHz, respectively. A total of 960
frequency channels is used, generated by polyphase filterbanks to
enable the correction of radio pulse dispersion in the interstellar
medium. Spectra are sampled every 65.4762\,$\mu$s.

The PALFA survey setup (high observation frequency, large number of
filterbank channels, fast-sampling spectrometers) is chosen to
maximize the chances of discovering MSPs at large distances within the
Galactic plane, where previous surveys have had little to no
sensitivity to MSPs. At the same time, the detection of these objects
has been difficult because of their high values of dispersion measure
(DM), which induce high dispersive smearing per channel. The narrow
channels used in the PALFA survey address this issue.  Finding many
MSPs is the highest priority in this survey because of their wide
range of astrophysical applications: testing Einstein's theory of
general relativity and alternative theories of gravity
\citep{2013Sci...340..448A}, measuring NS masses
\citep{2010Natur.467.1081D}, which can strongly constrain the equation
of state of dense matter, finding suitable sources for pulsar timing
arrays \citep{2010CQGra..27h4013H}, which will be used for detection
of very low-frequency gravitational waves, improved estimates of the
Galactic MSP population \citep{2014ApJ...787..137S}, and -- in the
case of \psr{} and similar systems -- a better understanding of
stellar and NS formation and evolution. PALFA discoveries are proof of
the power of the survey to discover MSPs at high DMs
\citep{2008Sci...320.1309C, 2010Sci...329.1305K, 2011ApJ...732L...1K,
  2012ApJ...757...90C, 2012ApJ...757...89D}.

Since its first observations in 2004, the PALFA Consortium has been
surveying the part of the sky close to the Galactic plane
($\left|b\right|\leq5^\circ$) that is visible to Arecibo Observatory,
i.e., declinations $0^\circ\leq\delta\leq37^\circ$. The complete
survey of this sky area will require about 330,000 separate Arecibo
beams (equivalent to $\sim$ 47,000 pointings of the seven-beam ALFA
receiver).

Data from the PALFA survey are analyzed by the \EAH{} pipeline briefly
described below and also by an independent pipeline operating on a
supercomputer at McGill University using the \texttt{PRESTO} software
package\footnote{\url{https://github.com/scottransom/presto}} (Lazarus
et al., in prep.).

\subsection{Einstein@Home}\label{subsec:pipeline}
\EAH{}\footnote{\url{http://einsteinathome.org}} is a distributed
volunteer computing project
\citep{Anderson:2006:DRS:1188455.1188586}. Members of the public
donate otherwise unused compute cycles on their home and/or office
PCs, and Android devices to the project to enable blind searches for
unknown NS.

\EAH{} is one of the largest distributed computing projects. In the
last ten years since its launch, more than 390,000~volunteers have
contributed to the project. On average, about 46,000~different
volunteers donate computing time each week on roughly
105,000~different
hosts\footnote{\url{http://einsteinathome.org/server_status.html} as
  of early April 2015}. The sustained computing power provided by
these volunteers is currently of order 1.75\,PFlop\,s$^{-1}$, which is
$\sim$5\% the computing power of the world's fastest supercomputer

\EAH{} analyzes data from the LIGO gravitational-wave detectors, the
\textit{Fermi Gamma-ray Space Telescope}, and large radio telescopes
such as the Arecibo Observatory and the Parkes Radio
Telescope. Observational data are stored and prepared for \EAH{}
processing on the Atlas computer cluster at the Albert Einstein
Institute in Hannover, Germany \citep{AtlasRef}. Volunteers' computers
download the data along with scientific software from dedicated
servers in Hannover and Milwaukee, and run it automatically. The
central coordination and management of the computing is handled by the
Berkeley Open Infrastructure for Network Computing (BOINC;
\citet{Anderson:2006:DRS:1188455.1188586}).

The \EAH{} PALFA analysis pipeline consists of three main steps. 1)
Observational data are de-dispersed at $\sim$4,000 trial values to
mitigate the radio pulse dispersion from the signal passing through
the interstellar medium. Strong burst-like RFI is masked and periodic
RFI is identified and replaced by random noise. Step 2) is done on the
computers of the general public, attached to the project: each
resulting de-dispersed time series is analyzed for periodic radio
pulsar signals using Fourier methods. The \EAH{} pipeline searches for
radio pulsars in compact binary systems with orbital periods as short
as 11\,min. The required orbital demodulation to remove the Doppler
effect from binary motion is done in the time domain. The demodulation
is repeated for $\sim$ 7,000 different orbital configurations to cover
a wide range of possible physical orbital parameters. For each of
these, the Fourier analysis is repeated after the time-domain
demodulation. The 100 statistically most significant candidates from
each de-dispersed time series are stored and sent back to the project
servers in Hannover. 3) The resulting $\sim$ 400,000 candidates for
each beam are sifted and remaining candidates folded using the raw
data and graded by machine learning methods
\citep{2014ApJ...781..117Z}.  Per-beam overview plots are also
visually inspected to identify (of order a few) promising candidates
in each beam.

A full description of the \EAH{} radio pulsar search pipeline employed
to discover \psr{} is beyond the scope of this publication and is
available in \citet{2013ApJ...773...91A}.  The sifting techniques used
to reduce the number of relevant candidates are described in
\citet{benthesis, 2013ApJ...774...93K}

To date, the project has discovered 51 neutron stars through their
radio emission.  As part of the PALFA collaboration, \EAH{} has
discovered a total of 27 radio pulsars\footnote{All discoveries are
  available online at
  \url{http://einsteinathome.org/radiopulsar/html/rediscovery_page/rediscoveries.html}
  and
  \url{http://einsteinathome.org/radiopulsar/html/BRP4_discoveries/}},
including the fastest spinning disrupted recycled pulsar (i.e., a
pulsar ejected from a binary system due to the companion star's
supernova explosion; \citealp{2010Sci...329.1305K}), and a
relativistic intermediate-mass binary pulsar
\citep{2011ApJ...732L...1K, 2014MNRAS.437.1485L}.  The \EAH{} search
for radio pulsars in archival data from the Parkes Radio Telescope has
found 24 pulsars missed by several previous re-analyses of this data
set \citep{2013ApJ...774...93K}.

\EAH{} has to date also discovered four gamma-ray pulsars in
\textit{Fermi} data \citep{2013ApJ...779L..11P}. These ongoing
searches use highly efficient data analysis methods initially
conceived for gravitational-wave data analysis
\citep{2009PhRvL.103r1102P, 2010PhRvD..82d2002P, 2011PhRvD..83l2003P,
  2012ApJ...744..105P}.

The detection of continuous gravitational waves from rotating neutron
stars in data from ground-based interferometric detectors is the main,
long-term goal of \EAH{}. The project has provided the most stringent
upper limits for gravitational waves from rapidly rotating neutron
stars in blind searches to date \citep{2009PhRvD..79b2001A,
  2009PhRvD..80d2003A, 2013PhRvD..87d2001A}.

\subsection{Discovery of \psr{}}\label{subsec:discovery}
\psr{} was discovered by the \EAH{} PALFA pipeline on 2011 October 4
with a statistical significance $\mathcal S=120.8$ (negative decadic
logarithm of the false-alarm probability in Gaussian noise) in a PALFA
survey observation from 2009 April 4. The pulsar was found at a spin
period of $4.3$\,ms and a DM of $142$\,pc\,cm$^{-3}$. The signal's
celestial nature was confirmed the same day with a second observation
by the Arecibo telescope.

\section{Observations and Data Analysis}\label{sec:obsanddataan}

\subsection{Observations}\label{subsec:obs}

Following its discovery, \psr{} was observed during the PALFA survey
observing sessions and with dedicated observations by the 76-m Lovell
telescope at Jodrell Bank.

The observations with the Arecibo telescope employed the usual PALFA
survey set up as described in Section~\ref{subsec:palfasurvey}, using
only the central beam of ALFA. The durations were between 4.5\,mins
and 15\,mins. In total 24 PALFA survey observations of \psr{} were
carried out from 2011 October 5 until 2012 May 5.

Between 2012 September 21 and 2012 December 31, nine dedicated timing
observations of \psr{} were carried out as part of a PALFA millisecond
pulsar timing campaign. Observations used the L-wide receiver and the
Puerto-Rican Ultimate Pulsar Processing Instrument (PUPPI) backend.  A
frequency band of 800\,MHz in 2048 channels, centered on a frequency
of 1380\,MHz was observed in all sessions. The usable bandwidth is
limited by the L-wide receiver which delivers an observation bandwidth
of 700\,MHz from 1.1\,GHz to 1.8\,GHz. Observation times were 10\,mins
or 15\,mins with a sampling time of 40.96\,$\mu$s.

The observations with the 76-m Lovell telescope used a
cryogenically-cooled dual-polarization receiver at a central frequency
of about 1520\,MHz. A 512-MHz band was sampled at 8-bits resolution
and processed using a digital filter bank into 2048 0.25-MHz frequency
channels.  After radio frequency interference excision, approximately
384\,MHz of usable bandwidth remained. The data were folded into 1024
pulse phase bins and were de-dispersed, generating an average profile
for each 10-second sub-integration. Observations of mostly 30\,mins
duration were made approximately every 10\,days and the data set
includes 61 TOAs from the MJD period 55851 to 56272.

\subsection{Data reduction}

Observational data were processed offline for radio frequency
interference (RFI) mitigation, and to obtain times-of-arrival (TOAs).

The separate sub-band data files obtained with the PALFA Mock
spectrometers were merged into one file in PSRFITS format for each
observation. PUPPI data were written in separate PSRFITS files, each
covering adjacent observation time stretches.

For RFI mitigation, the \texttt{rfifind} program from the
\texttt{PRESTO} pulsar processing suite was used to generate an RFI
mask for each observation.

\begin{figure}
\includegraphics[width=\columnwidth]{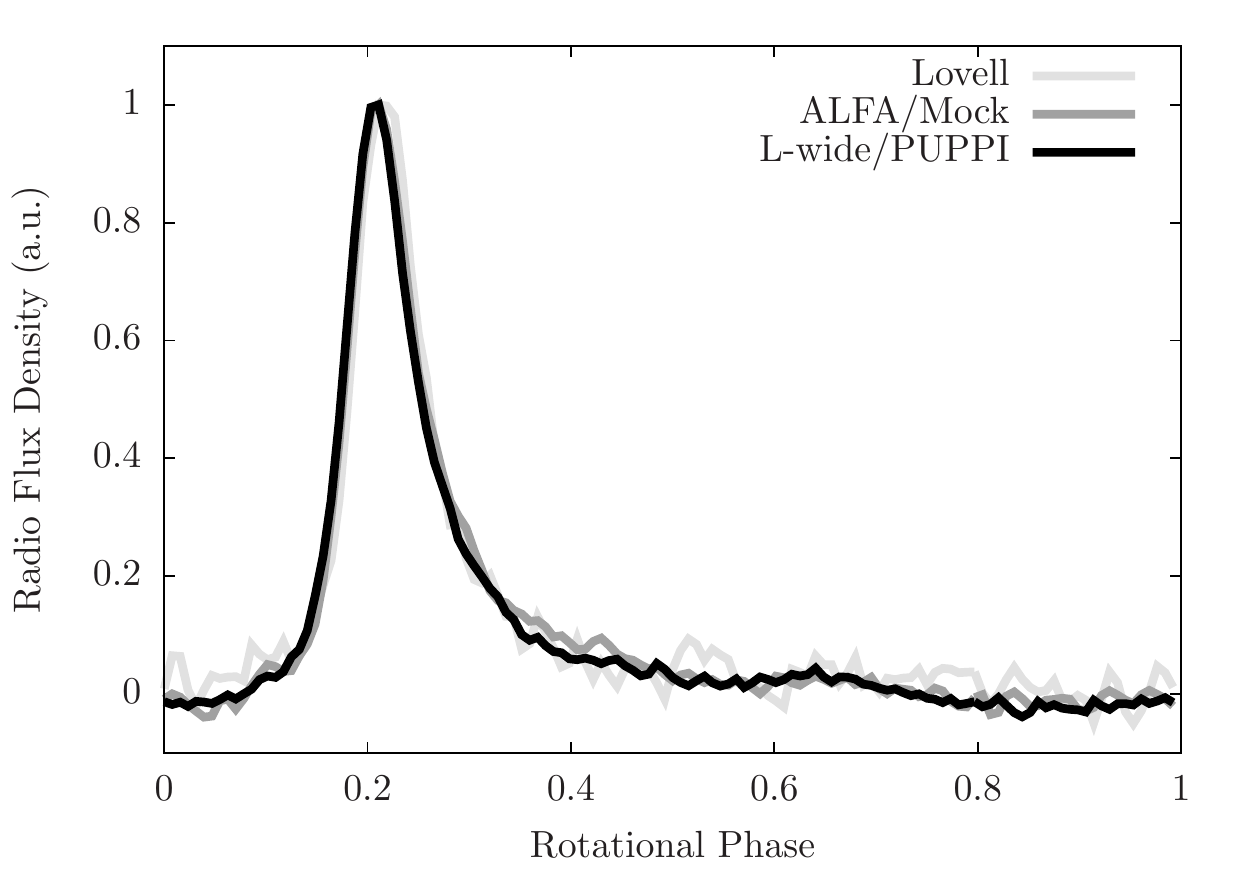}
\caption{The averaged radio flux density pulse profiles from the data
  collected with Arecibo Observatory using the Mock spectrometers and
  the ALFA receiver (dark-grey line) and with the PUPPI backend and
  the L-wide receiver (black line), respectively.  The averaged radio
  flux density pulse profile from observations at the Lovell Telescope
  in Jodrell Bank is shown (light-grey line). The spin period of the
  pulsar is $4.3$\,ms. These profiles were used to obtain the
  TOAs. All pulse profiles have been normalized and aligned at their
  maximum. The period-averaged flux density at 1.4\,GHz is
  $S_\text{1400}= 119\pm13$\,$\mu$Jy.}
\label{fig:profiles}
\end{figure}

Each observation was folded at the appropriate dispersion measure and
topocentric spin period with the \texttt{prepfold} program from
\texttt{PRESTO}.  Folded data files were used to create summed pulse
profiles from all ALFA/Mock spectrometer observations and all
L-wide/PUPPI observations, respectively.  Fig.~\ref{fig:profiles}
shows these pulse profiles, averaged over 25 observations with the
Mock spectrometers and over ten observations with the PUPPI backend,
respectively. The full width at half maximum duty cycle is 8\%. This
corresponds to a width of the pulse of $w_{50} = 0.34$\,ms. The
dispersive delay across a single frequency channel is 0.18\,ms for the
given DM, frequency resolution and central frequency. An exponential
scattering tail is apparent in the folded pulse profile. Centered on a
rotational phase of 0.65 a small additional pulse component appears in
both the ALFA/Mock and L-wide/PUPPI observations.

We calibrate the profile using the radiometer equation (e.g.,
\citealp{2012hpa..book.....L}) \beq \sigma =
\frac{S_\text{sys}}{\sqrt{2\Delta_f T_\text{obs} / N}} \eeq to predict
the observing system's noise level $\sigma$. We assume a system
equivalent flux density $S_\text{sys} = 2.4$\,Jy for the L-wide
receiver \footnote{Table~3 in
  \url{http://www.naic.edu/~astro/guide/guide.pdf}}. The combined
observation bandwidth of L-wide with the PUPPI backend is $\Delta_f =
700$\,MHz. We used a pulse profile summed from nine individual
observations, six of which with 9.8\,mins duration and three with
14.9\,mins duration.  The summed pulse profile was folded into $N=128$
bins; then the expected off-pulse noise standard deviation used for
calibration is 9.2\,$\mu$Jy. Scaling the off-pulse noise level of the
observations (between pulse phases 0.73 and 1.07 in
Fig.~\ref{fig:profiles}), the resulting estimated period-averaged flux
density of the pulsar at 1.4\,GHz is $S_\text{1400}=119\pm13\,\mu$Jy.

The \texttt{get\_TOAs.py} routine from \texttt{PRESTO} was used to
extract TOAs by employing the sums of pulse profiles as templates. A
single TOA was extracted for each ALFA/Mock spectrometer and
L-wide/PUPPI observation, respectively.  A total of 33 TOAs was
extracted from the data obtained with the Arecibo telescope, 24 of
which with the Mock spectrometer, the remaining 9 with the PUPPI
backend.

The PSRCHIVE\footnote{\url{http://psrchive.sourceforge.net}} package
by \cite{2004PASA...21..302H} was used for the data inspection,
interference removal and arrival time determination in the Lovell
Telescope data. A single TOA was generated for each of the
observations using a high signal-to-noise template formed from the
observations.

\begin{figure}
\includegraphics[width=\columnwidth]{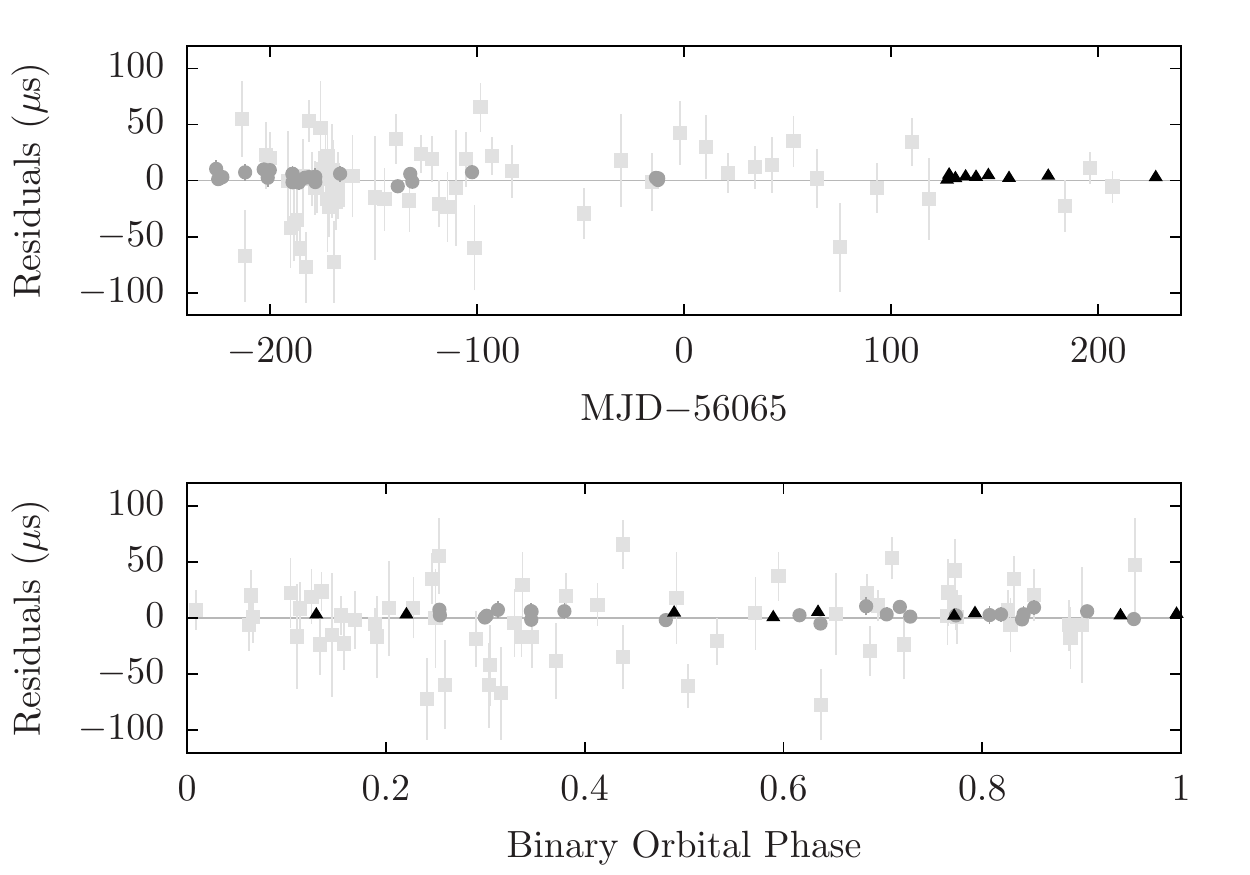}
\caption{The TOA residuals of our timing model as a function of MJD
  (upper panel) and binary orbital phase (lower panel). Light-grey
  squares show TOA residuals from observations with the 76-m Lovell
  telescope, dark-gray circles residuals from observations with the
  Arecibo telescope using the ALFA receiver and the Mock
  spectrometers, and black triangles observations with the Arecibo
  telescope using the L-wide receiver and the PUPPI backend. The TOA
  residual distribution does not exhibit any clear trends.}
\label{fig:residuals}
\end{figure}

The timing analysis used the \ttwo{} software
\citep{2006MNRAS.369..655H}. Our timing solution uses the BT orbital
model \citep{1976ApJ...205..580B, 2006MNRAS.372.1549E}. The coherent
solution fits 94 TOAs obtained between MJDs 55839 and 56293, covering
a baseline of 454\,days. As visible in Fig.~\ref{fig:residuals}, the
timing model accurately predicts TOAs. There are no clear trends in
the TOA residuals as a function of time, nor as a function of binary
orbital phase. Constant time offsets between the three TOA data sets
have been fitted with \ttwo{}.

The weighted RMS of the timing residuals is 4.7\,$\mu$s when including
TOAs from both observatories with all receivers and backends. The most
precise TOAs are obtained with the PUPPI backend and the L-wide
receiver at Arecibo. For an observation time of 15\,mins, the TOAs
have an average precision of 1.3\,$\mu$s. This makes \psr{} an
interesting pulsar to be used in Pulsar Timing Arrays
\citep{2010CQGra..27h4013H}.

\section{Timing Results}\label{sec:timres}
Astrometric, spin, and binary parameters, including the relativistic
periastron advance, were determined from the timing analysis of the
TOAs described in the previous section.
Table~\ref{tab:timingsolution} shows the parameters of our timing
solution obtained with \ttwo{}.

\begin{table}
\caption{Fitted and derived parameters for \psr{}.}
\label{tab:timingsolution}
\begin{tabular}{lc}
        \hline
        Parameter & Value$^{a}$\\
        \hline
        \multicolumn{2}{c}{\textit{General Information}} \\
        \hline
        MJD Range & 55839.0---56292.7 \\
        Number of TOAs & 94 \\
        Weighted RMS of Timing Residuals ($\mu$s) & 4.7 \\
        Reduced-$\chi^2$ value & 1.2 \\
        MJD of Period Determination & 55839 \\
        Binary Model Used & BT \\
        \hline
        \multicolumn{2}{c}{\textit{Fitted Parameters}} \\
        \hline
        R.A., $\alpha$ (J2000) & 19:50:45.06390(10) \\
        Dec., $\delta$ (J2000) & +24:14:56.9638(11) \\
        MJD of Position Determination & 55839 \\
        Spin Frequency, $\nu$ (Hz) & 232.30014862462(14) \\
        Spin Frequency Derivative, $\dot{\nu}$ ($\times 10^{-15}$ Hz\,s$^{-1}$) & $-$1.020(6) \\
        Dispersion Measure, DM (pc~cm$^{-3}$) & 142.089(18)$^{b}$ \\
        MJD of DM Determination & 55839 \\
        Projected Semi-Major Axis, $a_1\,\sin\left(i\right)$ (lt-s) & 14.2199738(11) \\
        Orbital Period, $P_\text{orb}$ (days) & 22.1913727(10) \\
        Epoch of periastron, $T_0$ (MJD) & 55846.0226219(15) \\
        Longitude of periastron, $\omega_0$ (deg) & 274.4155(3) \\ 
        Periastron advance, $\dot{\omega}$ (deg\,yr$^{-1}$) & 0.0020(3) \\ 
        Orbital eccentricity, $e$ & 0.07981158(12) \\ 
        \hline
        \multicolumn{2}{c}{\textit{Derived Parameters}} \\
        \hline
        Spin Period, $P$ (ms) & 4.304775549739(2) \\
        Spin Period Derivative, $\dot P$ ($\times 10^{-20}$) & 1.8900(10) \\
	Intrinsic $\dot P$, $\dot P_\text{int}$ ($\times 10^{-20}$) & $\le$2.1166(10)$^{c}$ \\
        Galactic longitude, $l$ (deg) & 61.10 \\
        Galactic latitude, $b$ (deg) & $-$1.17 \\
        Distance, $d$ (NE2001, kpc) & 5.5$^{+0.8}_{-0.8}$ \\
        Mass Function, $f_1$ ($M_{\odot}$) & 0.00626918(2) \\
        Minimum companion mass $M_2$ ($M_\odot$) & 0.253 \\
        Total Mass, $M$ ($M_{\odot}$) & 2.3(4) \\
        Characteristic Age, $\tau_c = P/(2\dot{P})$ (Gyr) & $\ge$3.2$^{c}$ \\
        Surface Magnetic Field Strength, $B_\text{S}$ ($\times 10^{8}$\,G) & $\le$3.1$^{c}$ \\
        Spin-down Luminosity, $\dot{E}$ ($\times 10^{34}$\,ergs/s) & $\le$ 1.0$^{c}$ \\
        \hline
\end{tabular}
$^{a}$ The numbers in parentheses are the 1-$\sigma$, \ttwo{}-reported
uncertainties on the last digit(s). Uncertainties in the pulsar
distance inferred from NE2001 are assuming a 20\% uncertainty in DM to
account for model uncertainties.\\ $^{b}$ The solution here is based
on single-frequency TOAs per Arecibo and Jodrell Bank observations,
respectively. Fitting the DM is therefore degenerate with fitting the
JUMPs between the data sets. The DM and its uncertainty were therefore
fitted from TOAs obtained in two sub-bands for the Arecibo data
combined with the Jodrell Bank TOAs, and kept fixed for the solution
presented here.\\ $^{c}$ The observed period derivative has been
corrected for Galactic acceleration. We cannot correct for the
Shklovskii effect, and therefore the value of $\dot P_\text{int}$ is
an upper limit, see Section~\ref{sec:timres}. Its reported uncertainty
here does not include the uncertainty from the Galactic
model. $\tau_c$, $B_\text{S}$, and $\dot{E}$ have been inferred from
$\dot P_\text{int}$.
\end{table}

The pulsar's mass function is given by \beq
f_1=\frac{4\pi^2x_1^3}{T_\odot P_\text{orb}^2} = \frac{M_2^3
  \sin^3\left(i\right)}{\left(M_1 + M_2\right)^2}, \eeq where $x_1 =
a_1\sin\left(i\right)/c$ is the pulsar's projected orbital semi-major
axis in light-seconds, $T_\odot = G M_\odot / c^3 =
4.925490947$\,$\mu$s is the solar mass in time units, $P_\text{orb}$
is the orbital period, and $M_1$ and $M_2$ are the masses of the
pulsar and the companion in units of solar masses, respectively. We
find $f_1=0.00626918(2)\,M_\odot$, which indicates a low-mass
companion with a minimum mass $M_2\geq 0.253\,M_\odot$, which is
obtained for $i=90^\circ$ and a pulsar mass $M_1 = 1.35\,M_\odot$. The
median companion mass, assuming $i=60^\circ$, is $M_\text{2, med}=
0.297\,M_\odot$. Therefore, the binary companion is most likely a
helium WD.

Considering the system's mass function, its high spin frequency
$\nu=232.300$\,Hz and small spin period derivative, \psr{} is very
similar to the majority of comparable binary pulsar systems in the
Galactic field. The most probable nature for the system would
therefore be a fully recycled millisecond pulsar with a WD
companion. In this case, mass transfer from the WD progenitor to the
MSP in the system's past would have circularized the binary orbit,
spun up the pulsar, and damped its magnetic field, as described in
Section~\ref{sec:introduction}.

However, the orbital eccentricity $e=0.07981158(12)$ of \psr{} is
larger (by roughly two orders of magnitude or more) than that of all
millisecond pulsars with helium WD companions in the Galactic field
with the exception of the three recently discovered similar systems
(see Sections~\ref{sec:introduction} and \ref{sec:discussion}).

The theory of general relativity predicts an advance of the longitude
of periastron, depending on the total mass of the system, its
eccentricity and orbital period \citep{1981GReGr..13....1W}, given by
\beq \dot \omega
=3\left(\frac{P_\text{orb}}{2\pi}\right)^{-5/3}\frac{\left(T_\odot\left(M_1
  + M_2\right)\right)^{2/3}}{1-e^2}, \eeq where $P_\text{orb}$, $M_1$,
$M_2$ and $T_\odot $ are defined as above and $e$ is the orbital
eccentricity.

Our observations of \psr{} significantly detect the system's
periastron advance $\dot{\omega} = 0.0020(3)$\,deg\,yr$^{-1}$. This
corresponds to an inferred total mass of $M = M_1 + M_2 =
2.3(4)\,M_\odot$, where the uncertainty is dominated by the
uncertainty in $\dot \omega$.  This measurement is suggestive of a NS
significantly more massive than 1.35\,$M_\odot$, however, it is
currently not precise enough to infer the individual masses of the
pulsar and/or the companion star. We expect our ongoing observation
campaign to significantly improve the measurement of $\dot \omega$ and
therefore the estimate of the total mass.

Ongoing timing observations might also help to constrain or measure
the Shapiro delay in the binary system. Combined measurements of the
Shapiro delay and the periastron advance can be used to infer the
separate masses of both components \citep{2010MNRAS.409..199F}.

The observed spin period derivative $\dot P$ has been corrected for
Galactic acceleration \citep{1995ApJ...441..429N}, to obtain a value
of the spin period derivative of $\dot P_\text{corr} = 2.12 \times
10^{-20}$. This takes into account a Galactic acceleration model that
uses the radial velocity curve published in
\citet{2014ApJ...783..130R}.  However, since the proper motion
currently cannot be measured with useful accuracy, we cannot correct
for the contribution $\dot{P_\text{Shk}}$ to the period derivative
from the Shklovskii effect \citep{1970SvA....13..562S}, and therefore
we can only estimate an upper limit for the intrinsic
$\dot{P_\text{int}}$.

Both the intrinsic $\dot{P_\text{int}}$ and the Shklovskii effect
contribution $\dot{P_\text{Shk}}$ are always positive and fulfill
$\dot P_\text{int} + \dot{P_\text{Shk}} = \dot
P_\text{corr}$. Therefore, assuming $\dot{P_\text{Shk}} = 0$, we
obtain an upper limit to the intrinsic spin period derviative of $\dot
P_\text{int} \le 2.12 \times 10^{-20}$. On the other hand, assuming
$\dot{P_\text{int}} = 0$ we can constrain the total proper motion to
$\le 19$\,mas/yr.

From $\dot P_\text{int} = 2.12\times 10^{-20}$\,s\,s$^{-1}$ we infer a
lower limit for the characteristic age of $\tau_c \ge 3.2$\,Gyr and
upper limits for the the surface magnetic field strength $B_\text{S}
\le 3.1\times 10^8$\,G and spin-down luminosity $\dot{E} \le 1.0\times
10^{34}$\,ergs/s. These parameters are very similar to those of the
general MSP population.

\subsection{Counterparts at other Wavelengths}\label{subsec:counterparts}

Let us now consider whether \psr{} is observable at X-ray and
$\gamma$-ray wavelengths and whether the nature of its companion can
be established by way of optical/IR observations.

The field around \psr{} has not been previously targeted in a pointed
observation by any former or current X-ray mission. A 581-s exposure
from the \textit{ROSAT} all-sky survey is too shallow to yield any
useful constrains on the X-ray flux from the pulsar. Based on its
$\dot{E}$, \psr{} likely has a soft, thermal X-ray spectrum due to
heating of its magnetic polar caps, with $L_X\approx
10^{-4}-10^{-3}\dot{E}\sim10^{30-31}$ erg s$^{-1}$, as typically seen
in non-eclipsing MSPs
\citep{2006ApJ...638..951Z,2006ApJ...646.1104B}. At a distance of
5.5\,kpc and assuming an absorbing column of $N_{\rm
  H}\approx4\times10^{21}$\,cm$^{-2}$ \citep[based on the empirical
  DM-$N_{\rm H}$ relation from][]{2013ApJ...768...64H} a detection of
the pulsar with \textit{Chandra} or \textit{XMM-Newton} would require
impractically long exposures.

There is no \textit{Fermi} LAT Third Point Source Catalog
\citep[3FGL;][]{2015arXiv150102003T} object in the vicinity of
\psr{}'s position.  A visual inspection of the \textit{Fermi} LAT data
from the beginning of the mission up to 2015 January 19 shows no
excess emission at the pulsar position.  This is expected given that
at the Galactic latitude of \psr{} ($b=-1.1^{\circ}$), the diffuse
$\gamma$-ray background is overwhelming. In addition, the commonly
used $\gamma$-ray ``detectability'' metric for pulsars,
$\sqrt{\dot{E}}/D^2$ is $3\times10^{15}$ erg$^{1/2}$ s$^{-1/2}$
kpc$^{-2}$ for \psr{}, substantially lower compared to MSPs detected
by \textit{Fermi} LAT \citep[see, e.g., Figure 15
  in][]{2013ApJS..208...17A} so it is not expected to be a bright
$\gamma$-ray source.

In principle, the nature of the companion of \psr{} can be established
by way of optical/IR observations.  If the secondary is a main
sequence star -- in this case \psr{} would belong to the same class as
of systems as PSR~J1903$+$0327 -- it is likely to be a M3.5 red dwarf
based on the mass measurement from radio timing
($\sim$0.25\,M$_{\odot}$). Data from the Digitized Sky Survey reveal
no optical counterpart at the radio position of the MSP, likely
because it is too distant and extincted.  Indeed, using the relations
by \citet{1995A&A...293..889P} and \citet{2013ApJ...768...64H}
combined with the pulsar's dispersion measure, we estimate the visual
extinction to the target to be $A_V\approx 2.3$.  Taking the absolute
magnitude of a M3.5V star, $M_V \approx 11.5$
\citep{2002AJ....123.2002H}, and scaling with the dispersion measure
derived distance ($\approx$5.5 kpc) yields an expected apparent
magnitude of $V \approx 27.5$ including extinction.

If we consider the intrinsic colors of an M3.5V star based on 2MASS
data \citep{2011AJ....142..138L}, and scale the extinction with
wavelength \citep[based on][]{1989ApJ...345..245C}, we obtain apparent
magnitudes in the near-IR of $J\approx23.9$, $H\approx23.1$ and,
$K\approx22.6$. Based on this, it is evident that near-IR imaging
observations provide the better means of establishing the nature of
the companion, which in turn helps constrain the evolutionary history
of \psr{}. Existing near-IR data from the 2MASS all-sky survey
\citep{2006AJ....131.1163S} show no counterpart at the pulsar
position, although this provides no meaningful constraints on the
\psr{} companion since the limiting magnitude is only $\sim$16. Thus,
deeper near-IR observations are required.
 
If the companion star is a He WD, its absolute magnitude is $15 > M_V
> 10$ \citep{2006ces..book.....K}. The extinction given above and the
DM-inferred pulsar distance yield an apparent magnitude of $31 \gtrsim
m_V \gtrsim 26$. Detection in the visual band in imaging observations
with 8-meter class telescopes requires apparent magnitudes of $m_V<24$
\citep{2015MNRAS.446.4019B}.  Therefore, a detection or spectroscopic
identification of an He WD companion in the optical band is currently
impossible.

\section{Discussion}\label{sec:discussion}

\psr{} has an unusual combination of small pulsar spin period and high
orbital eccentricity.  The small spin period and small spin period
derivative (and inferred surface magnetic field) point to a long
recycling episode in the system's past, in which matter from the
companion star accreted onto the pulsar, spun it up and reduced its
magnetic field. Yet, the high orbital eccentricity appears at odds
with this picture because long recycling episodes are believed to
circularize the binary orbits to eccentricities in the range
$e=10^{-7} - 10^{-3}$ \citep{1992RSPTA.341...39P}.

Thus, these systems must form through different channels and/or
experience some mechanism that increases the orbital eccentricity
after the pulsar is recycled. We will discuss four proposed channels:
1) Formation in a hierarchical triple system which became unstable and
ejected one of its members; 2) Perturbations in regions with high
stellar density (globular clusters); 3) The rotationally delayed
accretion-induced collapse of a super-Chandrasekhar white dwarf; 4)
Dynamical interaction of the binary with a circumbinary disk.

\begin{table}
\begin{center}
\caption{Physical parameters of the five known eccentric
  MSPs. PSR~J1903$+$0327 \citep{2008Sci...320.1309C} was most likely
  formed from a triple system. PSR~J1618$-$3921 (data reproduced from
  \citealp{2001ApJ...553..801E}, \citealp{2010NewAR..54...80B}, and
  Bailes -- private communication), two recently discovered MSPs
  \citep[data from][]{2014MNRAS.438L..86F,
    2013MNRAS.435.2234B,2013ApJ...775...51D} and \psr{} (from this
  publication) have very similar orbital and spin characteristics,
  suggesting a common formation channel. The height above the Galactic
  plane, $z$, was calculated from the DM-inferred distance and the
  Galactic latitude.}
\begin{tabular}{lccccc}
\hline
\hline
Pulsar & $P$ & $P_\text{orb}$ & $M_{\rm 2, med}$ & $e$ & $|z|$ \\
\hline
\hline
\noalign{\smallskip} 
	PSR~J1903$+$0327	& 2.1\,ms	& 95.2\,days & $1.1\,M_\odot$ & 0.44 & 0.1\,kpc\\
	\hline
        PSR~J2234$+$06     &  3.6\,ms & 32\,days & $0.23\,M_\odot$ & 0.13 & 0.4\,kpc\\
        PSR~J1946$+$3417 &  3.2\,ms & 27\,days & $0.24\,M_\odot$ & 0.14 & 0.5\,kpc\\
        PSR~J1618$-$3921 & 12.0\,ms & 22.8\,days & $0.20\,M_\odot$ & 0.027 & 0.6\,kpc\\
        \textbf{\psr{}} &  \textbf{4.3\,ms} & \textbf{22.2\,days} & \textbf{0.30\,$M_{\odot}$} & \textbf{0.08} & \textbf{0.1\,kpc} \\
\noalign{\smallskip} 
\end{tabular}
\label{tab:systems}
\end{center}
\end{table}

Table~\ref{tab:systems} compares our discovery with the four known
eccentric MSPs in the Galactic field.

The similarity of the spin and orbital parameters of \psr{} to those
of PSR~J2234$+$06 and PSR~J1946$+$3417 (and also, to a lesser extent,
to PSR~J1618$-$3921) is striking, and is further evidence for the
existence of a common formation mechanism that is clearly distinct
from that of PSR~J1903+0327 (which, as mentioned before, likely
resulted from the chaotic disruption of a triple system, as indicated
by its more massive main-sequence companion, much larger eccentricity
and orbital period). PSR~J1618$-$3921 differs from the aforementioned
pulsars by a longer spin period and less eccentric orbit, however the
orbital period is similar to that of \psr{}, supporting the idea that
something anomalous happens within this range of orbital periods.

All pulsars in Table~\ref{tab:systems} have small distances from the
Galactic plane $|z|\lesssim0.6$\,kpc. If the systems were produced by
2) exchange interactions in globular clusters, their distances to the
Galactic plane are expected to be large, similar to the values for the
globular cluster population \citep{1996AJ....112.1487H}.  Therefore,
it is unlikely that any of the pulsars originated in a globular
cluster. Future observations of \psr{} might be used to constrain its
proper motion and further elucidate the issue.

As described in Section~\ref{sec:introduction},
\citet{2014MNRAS.438L..86F} recently have proposed the formation
mechanism 3) for fully recycled MSPs in eccentric orbits. Since \psr{}
is such a system, let us now consider whether it could have also
formed by the same mechanism.

At the center of their model is an accretion-induced collapse of a
super-Chandrasekhar mass white dwarf in a close binary. The WD only
collapses after the end of the accretion episode, and after it has
lost sufficient spin angular momentum, directly producing an MSP in an
eccentric orbit.  Based on their simulations,
\citet{2014MNRAS.438L..86F} provide a list of predictions for MSP
systems formed through the rotationally delayed AIC channel. At the
current time, \psr{} is consistent with these predictions as detailed
in the following.

From Fig.~3 in \citet{2014MNRAS.438L..86F}, and assuming a range of
pre-AIC WD masses between 1.37 and 1.48\,$M_\odot$, isotropically
directed kicks, and initial orbital periods between 15 and 30\,days,
the kick velocity must be around 2\,km\,s$^{-1}$ to yield orbital
eccentricities between 0.09 and 0.14. The orbital eccentricity of
\psr{} is just outside the lower end of that range (and that certainly
is the case for PSR~J1618$-$3921). This might suggest a slightly
larger kick magnitude and a slightly wider range of eccentricities to
be observed in these systems in the future, or a slightly smaller NS
binding energy than that assumed in that simulation.

The companion should be a He WD with a mass in the range $0.24 -
0.31\,M_\odot$, as predicted by the
$P_\text{orb}$-$\mathrm{M}_\text{WD}$ relation from Fig.~5 in
\citet{1999A&A...350..928T} for this orbital period of 22
days. Although we cannot currently determine the nature of the
companion of \psr{}, its median companion mass of 0.30\,$M_\odot$ is
certainly compatible with this expectation.

\citet{2014MNRAS.438L..86F} predict a pulsar mass of typically $1.22 -
1.31\,M_\odot$. Currently, we have not measured the total mass of the
system to a sufficiently high precision to test this prediction.

The system has low Galactic height, as expected from systems that
formed with very small kick velocity ($\lesssim 10$\,km/s) in an AIC
event. Indeed, the low Galactic heights of the MSPs in eccentric
orbits discovered to date suggests small peculiar space velocities
compared to the general MSP population.  This assumption might be
further tested with a future measurement of the proper motion of the
system.

The characteristics of \psr{} are also consistent with dynamical
interaction with a circumbinary disk \citep{2014ApJ...797L..24A}. The
circumbinary disk model makes two predictions which could be tested
for \psr{} with further observations: 1) As in
\cite{2014MNRAS.438L..86F}, the companions should also be He WDs with
masses given by \citet{1999A&A...350..928T}, which as we have seen
agrees with our current mass constraint for the companion of \psr{}.
2) The masses of the MSP and the peculiar space velocities should
closely resemble those of circular binary MSPs, unlike the
\textit{small} velocities ($\lesssim 10$\,km/s) predicted by the AIC
model. With our current timing data, we cannot constrain the proper
motion and can therefore neither confirm nor falsify this prediction.

Whatever the formation scenario for systems like \psr{} is, it has in
this case produced an object with an inferred surface magnetic field
comparable to the general MSP population. This does not falsify any of
the models discussed above, but might be useful constraint on other
possible formation scenarios.

The fact that the characteristic age ($\tau_c \ge 3.2$\,Gyr) is very
similar to that of the general MSP population can be used to formulate
a first rough estimate of the occurrence of these systems. It suggests
that the relative frequency of occurrence of \psr{}-like MSPs in the
population is comparable to the fraction of the currently known
systems in the currently known population. Assuming that PSRs
J1618$-$3921, J1946$+$3417, J1950$+$2414, and J2234$+$06 all formed
through the same mechanism and all have similar $\tau_c$, the relative
frequency of these systems is of order 4 of a total of 230 known
MSPs\footnote{\url{http://astro.phys.wvu.edu/GalacticMSPs/GalacticMSPs.txt}}
not associated with a globular cluster, roughly 2\% of the population.

\section{Conclusions and Future Work}\label{sec:conclusionfuture}

We have presented the discovery and initial timing of the
fully-recycled ($P=4.3$\,ms) MSP \psr{} in an eccentric ($e=0.08$)
orbit ($P_\text{orb} = 22$\,days) with a median companion mass of
0.3\,$M_\odot$. \psr{} is only the fifth system with a large
eccentricity to be discovered. Its spin and orbital parameters are
similar to those of three previously known systems with orbital
periods in the range of $\sim20-30$\,days, spin periods between 2\,ms
and 12\,ms, large orbital eccentricities ($0.03\lesssim e \lesssim
0.15$), and companions with masses $0.2\,M_\odot\lesssim m_\text{c}
\lesssim 0.3\,M_\odot$.

This combination of parameters cannot be explained in the standard
pulsar recycling scenario.  The existence of now four known similar
systems which probably have a helium WD companion suggests the
existence of a common formation channel leading to these unusual MSP
systems. The upper limit on the intrinsic spin period and the inferred
surface magnetic field are comparable to those of the general MSP
population.

We compared the properties of \psr{} with the predictions of four
proposed formation channels.  The initial evolution of the pulsar in a
hierarchical triple and the origin in an exchange encounter in an high
stellar density environment (e.g., globular cluster) are
unlikely. Although we cannot conclusively rule out or confirm
formation through a rotationally-delayed AIC event or interaction with
a circumbinary disk with the measurements presented here, future
observations might allow us to test a variety of predictions made by
the different possible formation models.

The pulsar is currently being observed with Arecibo Observatory, and
this will allow us to better constrain the pulsar and companion
masses, which is of highest importance to discriminate between the AIC
model and the circumbinary-disk model. The individual masses of the
pulsar and the companion can be obtained from combining observations
of the relativistic periastron advance with observations of the
Shapiro delay. Using the orthometric parametrization of the Shapiro
delay from \citet{2010MNRAS.409..199F} should provide the most precise
estimates of the individual masses. The uncertainties of the
periastron advance and the Shapiro delay in this parametrization are
less degenerate than in the standard parametrization. Therefore, even
if the measurement of the Shapiro delay has larger errors, a precise
measurement of the relativistic periastron advance will allow for
well-defined pulsar mass determination, see, e.g.,
\citet{2012ApJ...745..109L}. The ongoing observations will improve the
relativistic periastron advance measurement and might also lead to a
detection of the Shapiro delay.

Radio observations with longer time baselines will provide upper
limits on or a measurement of the proper motion of the pulsar, and
therefore of its peculiar space velocity. The AIC model predicts small
values ($\lesssim 10$\,km/s), while the circumbinary-disk model
predicts larger values.  Measuring the proper motion could therefore
permit discrimination between the two models.

Further information about the nature of the companion might be
obtained from observations at optical or infrared wavelengths, as
detailed in Section~\ref{subsec:counterparts}. Near-IR imaging
observations provide the best means of establishing the nature of the
companion, which in turn helps constrain the evolutionary history of
\psr{}.

Modern pulsar surveys like PALFA are probing the Galactic disk to
unprecedented depths and are likely to find other such systems. Having
a larger sample size will allow discriminating between the different
proposed models with greater confidence than is possible now.

\section*{Acknowledgements}\label{sec:acknowledgements}
We thank all \EAH{} volunteers, especially those whose computers found
\psr{} with the highest statistical significance\footnote{Where the
  real name is unknown or must remain confidential we give the \EAH{}
  user name and display it in single quotes.}: David Miller,
Cheltenham, Gloucestershire, UK and `georges01'.

The authors would like to thank the anonymous referee for the advice
and comments that helped to improve this manuscript.

This work was supported by the Max-Planck-Gesellschaft and by NSF
grants 1104902, 1105572, and 1148523.

The Arecibo Observatory is operated by SRI International under a
cooperative agreement with the National Science Foundation
(AST-1100968), and in alliance with Ana G. M{\'e}ndez-Universidad
Metropolitana, and the Universities Space Research Association.

I.H.S.\ and W.Z.\ acknowledge support from an NSERC Discovery Grant
and Discovery Accelerator Supplement and from CIfAR.

J.S.D. was supported by the Chief of Naval Research.

J.W.T.H.\ acknowledges funding from an NWO Vidi fellowship and ERC
Starting Grant ``DRAGNET'' (337062).

P.C.C.F.\ and L.G.S.\ gratefully acknowledge financial support by the
European Research Council for the ERC Starting Grant BEACON under
contract no.~279702.

V.M.K.\ acknowledges support from an NSERC Discovery Grant and
Accelerator Supplement, the FQRNT Centre de Recherche en Astrophysique
du Qu{\'e}bec, an R. Howard Webster Foundation Fellowship from the
Canadian Institute for Advanced Research (CIFAR), the Canada Research
Chairs Program and the Lorne Trottier Chair in Astrophysics and
Cosmology.

\bibliographystyle{yahapj}
\bibliography{paper}

\end{document}